\documentclass[12pt,a4paper]{article}
\usepackage{jheppub_kim}
 \usepackage{slashed}
\usepackage{longtable}
\usepackage{epsfig}
\usepackage{epstopdf}
\usepackage{amssymb,amsmath}
\usepackage{amsthm}
\usepackage{graphicx}
 \usepackage{graphics}

\usepackage[hang,nooneline,scriptsize]{subfigure}
\usepackage{subfigure}
\usepackage{array}
\begin {document}
\title{Small Black String Thermodynamics}
\author[a] {Jyotish Kumar,}
 \author [b,c,]{ Sudhaker Upadhyay\footnote{Corresponding author}\footnote{Visiting Associate,  IUCAA  Pune, Maharashtra-411007, India}} 
 \author[a]{and Himanshu Kumar Sudhanshu}

		\affiliation[a]{P.G. Department of Physics,   Magadh University, Bodh Gaya, Bihar  824234, India}
	 \affiliation[b]{Department of Physics, K. L. S. College, Magadh University, Nawada, Bihar 805110, India}
 
  \affiliation[c]{School of Physics, Damghan University, Damghan, 3671641167, Iran}

\emailAdd{Jyotishkumar7137@gmail.com }	
 \emailAdd{sudhakerupadhyay@gmail.com} \emailAdd{himanshu4u84@gmail.com}

\abstract{We consider a cylindrically symmetric solution for the field equations of Einstein-Hilbert action  with a negative cosmological constant in four dimensions. The small statistical fluctuation in the equilibrium thermodynamics of the black string solution is investigated.
The small  black string under the influence of small statistical fluctuation also follows the first law of thermodynamics.  The behaviour of
equation of states for black string changes significantly due to the fluctuation. 
The fluctuation causes instability to 
the small-sized black string only. Assuming the
black string is fluid, the compressibility of the black string is inversely proportional to the fluctuation parameter. 
  } 
\maketitle
 
 \section{Literature survey and motivation}
 
 The radiation and evaporation of the black holes \cite{1,2}  merge three areas of physics: quantum theory, general relativity and thermodynamics. Even though black holes cannot be seen directly, there is no doubt regarding their existence. 
 It is found that the entropy of black holes depends on the horizon's area. 
Nowadays, it is confirmed that the entropy of large black holes is  directly proportional to the horizon's area \cite{3,4}.  However, for
the entropy of small black holes, the small statistical fluctuations around the
equilibrium play an essential role \cite{5,6}.
This
is well-known that the leading-order correction to the entropy is always logarithmic due to small statistical
fluctuations around the equilibrium. 
Recently, thermodynamics of various small black holes under the influence of small statistical fluctuations have been investigated extensively \cite{7,8,9,10,11,12}.

{In general, the black hole solutions are axisymmetric solutions characterised by four parameters: mass, angular momentum, charge and
the cosmological constant. There are two possibilities for such axisymmetric solutions. One is a solution with spherical symmetry like the Schwarzschild solution, which has been studied extensively. The other is the cylindrical symmetric solution
(black or cosmic string), which is not studied extensively.  However, the cylindrical symmetry has been found very crucial in the various contexts of general relativity  like the static
solutions of Levi-Civita \cite{07,08} and Chazy-Curzon \cite{09,010}, and the stationary solutions
of Lewis \cite{011}. The cylindrical symmetry is used to study cosmic strings in astrophysics \cite{012}. The general cylindrically symmetric static solution of Einstein's equation  
with cosmological constant was found by Linet in Ref. \cite{013}, which was not a black hole solution.  Lamos found the first black hole solution with cylindrical symmetry in Ref. \cite{13}.
In general, black string and cylindrical black
hole are different, but they mimic each other in $ 4$ dimensions.}

 The spherically symmetric black holes were explored initially. The study of prolate gravitational collapse of cylindrical and
other similar objects discovered later  during the  hoop conjecture  formulation, which postulates 
that horizon formation is possible if the circumference of compressed mass is less than its Schwarzschild circumference in every direction.
This excluded the possibility of the formation of a cylindrical black hole.
 In fact,  
the hoop conjecture is valid only in the absence of a cosmological constant. However,  cylindrically symmetric black holes  exist in a negative cosmological constant.
 Recently,   the black strings have been constructed for 4-dimensional
Einstein gravity with a negative cosmological constant \cite{13}.
The thermodynamic properties of a small black string, where a change in entropy 
  due to thermal fluctuation playing a significant role is not studied yet.
This provides us with a motivation for the present investigation 
which will bridge the gap. 

In this paper, we mainly consider a system of a black string with a negative cosmological  constant and revisit   the thermodynamics of the black string where small statistical fluctuation for the entropy is 
not relevant, and eventually, entropy follows the area law.  
The small statistical fluctuation  plays a vital role in the thermodynamics of the small black strings, so we have considered the respective modification to the entropy of the black string.  Eventually,  the black string's whole thermodynamics gets corrected due to modified entropy. We further investigate the change in the behaviour of the thermal equation of states due to thermal fluctuations. Graphically, we observe that the additional terms of entropy of the black string are significant for small horizon black string. 
On the other hand, the fluctuation decreases the internal
 energy for the black string, which is significant for larger-sized black strings. 
The corrected Helmholtz free energy matches its uncorrected counterpart for a more extensive horizon radius.  Finally, the small statistical fluctuation causes the small black string radius instability. However, the black string remains stable without considering thermal fluctuation. 

 The paper is organised in the following manner.
 In section \ref{sec2}, we recapitulate the considered model describing black string and the cosmological constant. 
Following the black string mechanics, we discuss the introductory
thermodynamics of the system. In section,
\ref{sec3}, we investigate the impacts of small statistical 
fluctuation on the equilibrium thermodynamics of black string.
The stability and fluid compressibility of the black string is discussed in section \ref{sec4}. Here, the system of the black string is considered to follow fluid dynamics.
Finally, we conclude the discussion in section \ref{sec5}.
\section{Black string and its thermodynamics}\label{sec2}
The Einstein-Hilbert action  for the four-dimensional uncharged black string in the presence of cosmological constant (setting gravitational constant, $G=1$) is 
given by \cite{II1}
\begin{equation}
\label{Eq1}
I=\frac{1}{16\pi}\int d^4 x\sqrt{-g}(R+6\alpha^2),
\end{equation}
where $g$ is the determinant of the metric $g_{\mu \nu}$, $R$ is the Ricci scalar curvature, and $\alpha^2=-\frac{\Lambda}{3}>0$ denotes the negative cosmological constant.

Varying the Einstein-Hilbert action (\ref{Eq1}) with the metric  yields the field equation as
\begin{equation}
\label{Eq2}
 R_{\mu \nu}-\frac{1}{2}g_{\mu \nu}R=3 \alpha^2 g_{\mu \nu},
\end{equation}
where  $R_{\mu \nu}$ is the Ricci tensor.

The general metric solution for the Einstein equation (\ref{Eq1}) in the cylindrically symmetric and
 time-independent spacetime $(t, r, \phi ,z)$ is written by \cite{13,II1} \begin{equation}
\label{Eq3}
ds^2=-f(r)dt^2 +g(r) dr^2+r^2 d\phi^2+\alpha^2r^2 dz^2,
\end{equation}
here, $-\infty<t, 0\leq r<+\infty,0\leq \phi \leq 2\pi$ and $z<+\infty$. The metric function has the following form \cite{II1}:
\begin{equation}
\label{Eq4}
f(r)=g^{-1}(r)=\left(\alpha^2 r^2-\frac{4M}{\alpha r}\right),
\end{equation}
where $M$ is the integration constant related to the black string's Arnowitt-Deser-Misner (ADM) mass density. The singularity at $r=0$ is the cosmological one, and for a moment, we are not interested in such singularity conditions.

 The event horizon of the black string can be obtained from the root of the horizon condition, $f(r_+)=0$. 
 For positive $M$,   it has a horizon at $r_+=\frac{1}{\alpha}\left(4M\right)^\frac{1}{3}$
in the positive $r$ direction, it has a naked singularity  
in the negative $r$ direction. The situation is
reversed for negative $M$. This property is a new feature of such a black string.
 
  To study 
  thermodynamics of the black string solutions, we employ the Euclidean method \cite{0012,0013}.
  The requirement of the
absence of the conical singularity in the Euclidean spacetime
(\ref{Eq4}) causes the Euclidean time  must have a period
$T^{-1}$, which satisfies
\begin{equation}
 \label{Eq6}
 T=\left.\frac{f'(r)}{4\pi  }\right\vert_{r=r_+}=\frac{3\alpha^2 r_+}{4 \pi}.
 \end{equation}   
 which is just the Hawking temperature of the black
plane solutions (strings) calculated from surface
gravity. Here, prime ($'$) denotes the derivative concerning $r$.
  In terms of the mass of the uncharged black string, the Hawking temperature identifies to $T=(3\alpha/2 \pi)\left(M/2\right)^{1/3}$, varies as $M^{1/3}$, which is different from the regular Schwarzschild black holes. This implies that the black string has a different configuration from black holes due to changes in topological structures and, hence, it has a different thermodynamical configuration.
 
Since the  black string mimics a thermodynamical system    and, therefore, follows the first law of the thermodynamics
\cite{hen,gh}
\begin{equation}
dM=TdS_0,\label{xx}
\end{equation}
 where $S_0$ is the equilibrium entropy of the black string. 
 Here, we note that for the first law of black string thermodynamics, the work term (pressure and volume) is absent \cite{hen,gh}. However, one can have such a work term in extended phase space \cite{hen}.
  Using Eqs. (\ref{Eq6}) and (\ref{xx}), the first law of thermodynamics leads to 
 \begin{equation}
 \label{Eq7}
 S_0=\int \frac{dM}{T}=\frac{\pi \alpha r_{+}^2}{2}.
 \end{equation}
 In terms of the mass of the black string, entropy, $S_0=(\pi /2 \alpha)(4M)^{2/3}$, varies as $M^{2/3}$ and hence different configuration from the black hole.
 
From the  Area-law, $S_0=\sigma /4$ suggests that the area of the event horizon of the black string, $\sigma$, must have the following expression:
 \begin{equation}
 \sigma=2 \pi \alpha r_{+}^2=\frac{\pi}{2\alpha}\left(4M\right)^{\frac{2}{3}}.
 \end{equation}
 The thermodynamic volume of the black string can be calculated from the  area (entropy) relation as
\begin{equation}
\label{Eq9}
V= \int \sigma dr_+=\frac{2}{3}\pi \alpha r_+^3.
\end{equation}

We can now estimate the differences 
in the equilibrium thermodynamics due to the small statistical disturbances that become significant for quantum black bodies. 
\section{Non-equilibrium thermodynamics of   black string}\label{sec3}
It is well-known that the thermodynamics of the black bodies
gets logarithmic perturbation which plays a crucial role for the small black holes. Therefore, on this basis, we anticipate that the thermodynamics of black string will also follow the same trend.  
The nature of  the logarithmic correction to the entropy   due to statistical fluctuations at the equilibrium is given by \citep{III1, III2}
\begin{equation}
\label{EqIII1}
S=S_0-\beta \ln(S_0 T^2),
\end{equation}
where $\beta$ is a label parameter which is equal to $1/2$
when fluctuation exists and vanishes for the system at equilibrium. 

\begin{figure}[ht]
\centering
\includegraphics[width=300pt]{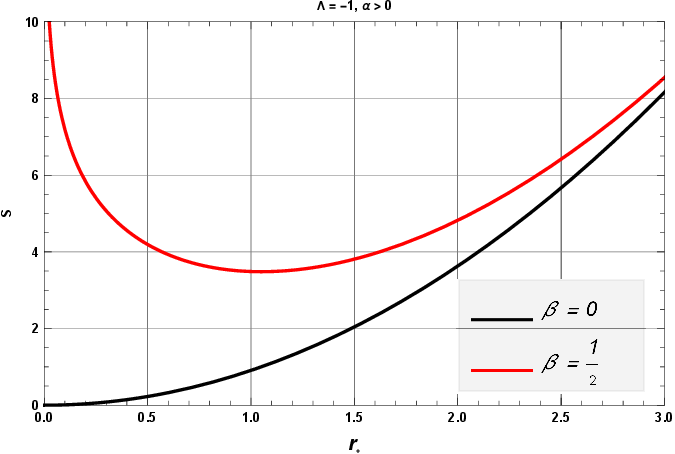} 
\caption{The plot of entropy $(S)$ with horizon radius $(r_+)$ of the black string.} 
\label{fig:Sc}
\end{figure}

For the given values of Hawking temperature (\ref{Eq6}) and  equilibrium entropy (\ref{Eq7}), the  entropy for the uncharged black string under thermal fluctuation reads
\begin{equation}
\label{EqIII2}
S=\frac{\pi \alpha r_{+}^2}{2}-\beta \ln\left(\frac{9 \alpha^5 r_{+}^4}{32 \pi}\right).
\end{equation}
To visualise the effects of  correction on the nature of 
entropy, we plot the equilibrium entropy and the corrected entropy of the black string due to thermal fluctuations in Fig.  \ref{fig:Sc}.
 The correction term does not show significant change for the larger radial circumference of the horizon of the black string. Still, for the smaller horizon radius, the corrected entropy of the black string is rather substantial and makes the entropy asymptotically prominent as the horizon ceases to point. 
The plot shows that entropy takes only a positive value in both cases without and with thermal corrections.

\begin{figure}[hbt]
\centering
\includegraphics[width=300pt]{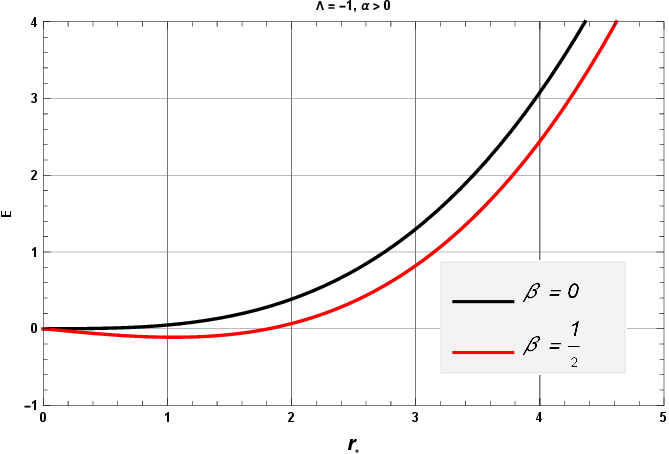} 
\caption{The plot of internal energy $(E)$ with horizon radius $(r_+)$ of the black string.} 
\label{fig:Sc1}
\end{figure}

Once we have Hawking temperature (\ref{Eq6}) and the corrected entropy (\ref{EqIII2}), the corrected internal energy of the black string, $E$, due to the thermal fluctuation at equilibrium is computed as
\begin{equation}
\label{III3}
E=\int TdS=\frac{\alpha^3 r_{+}^3}{4}-\beta \left(\frac{3 \alpha^2 r_+}{\pi}\right).
\end{equation}
Now, the behaviour of the above expression is depicted in Fig.
 \ref{fig:Sc1}, which studies the effects of fluctuation on the equilibrium internal energy. 
 The plot shows that the internal energy is an increasing function of the horizon radius.  Interestingly, the fluctuation decreases the value of internal
 energy for the black string, which is not rather significant for small black strings.  The internal energy of a small black string becomes negatively valued if we introduce thermal fluctuation to the system, which is quite an interesting result. {This suggests that the thermal fluctuation is taking energy from the system and dissipating away as the system is not doing any work.}

\begin{figure}[hbt]
\centering
\includegraphics[width=300pt]{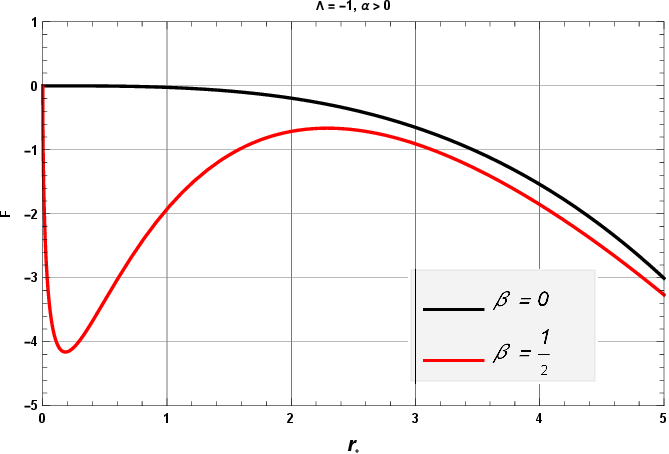} 
\caption{The plot of Helmholtz free energy $(F)$ with horizon radius $(r_+)$ of the black string.}
\label{fig:Sc2}
\end{figure}

The  Helmholtz free energy a for non-equilibrium black string can be obtained from the relation, $F=-\int S dT$. Utilizing the Hawking temperature (\ref{Eq6}) and the entropy (\ref{EqIII2}),
this leads to 
\begin{equation}
\label{III4}
F=  -\frac{\alpha^3 r_{+}^3}{8}+\frac{\beta \alpha^2 r_+}{4 \pi}\left[\ln \left(\frac{9\alpha^5 r_{+}^4}{32 \pi}\right)^3 -6\right].
\end{equation}
Now,   the behaviour of expression (\ref{III4}) is depicted in Fig. \ref{fig:Sc2}.
 Here, we observe that Helmholtz's free energy of the small-sized black string shows inverse power-law behaviour. However, it matches the equilibrium counterpart
for a more extensive horizon radius. A negative value of Helmholtz's free energy may indicate that the final energy of the black string is smaller than its original energy. {This result is also in agreement with the internal energy.} 

\begin{figure}[hbt]
\centering
\includegraphics[width=300pt]{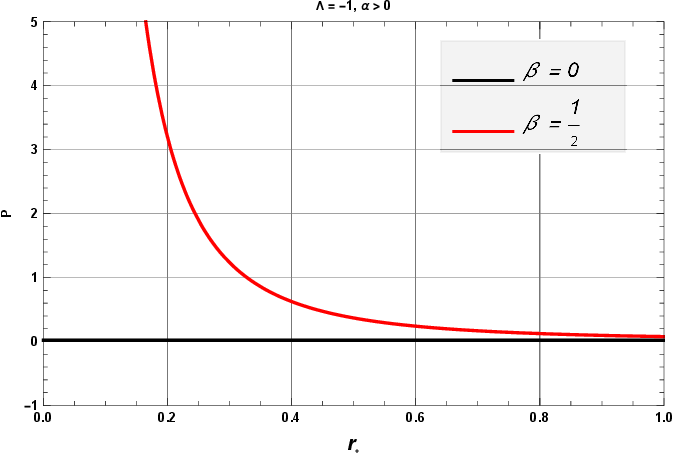} 
\caption{The plot of thermodynamic pressure $(P)$ with horizon radius $(r_+)$ of the black string.} 
\label{fig:Sc3}
\end{figure}

For the given Helmholtz free energy (\ref{III4}) and volume (\ref{Eq9}),
the  thermodynamic pressure of the black string is calculated in the extended phase space (by identifying the negative cosmological constant with the equilibrium pressure) with the help of relation, $P = -\left(\frac{\partial F}{\partial V}\right)$, as 
\begin{equation}
\label{III5}
P =\frac{3 \alpha^2}{16 \pi}+\frac{3 \beta \alpha}{8 \pi^2 r_{+}^2}\ln \left(\frac{32 \pi}{9 \alpha^5 r_{+}^4}\right).
\end{equation}
Now, we plot the pressure (\ref{III5}) concerning the horizon radius in Fig. \ref{fig:Sc3}.
 The pressure due to thermal fluctuation is asymptotically large for small black string. This means that thermal fluctuation for the system of small black string induces more stress on the system.  
However, the system has constant pressure  for large black string
for both the equilibrium and non-equilibrium conditions.

\begin{figure}[hbt]
\centering
\includegraphics[width=300pt]{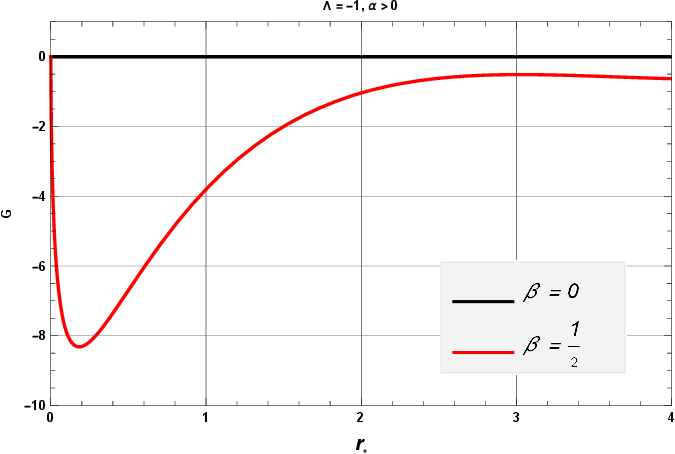} 
\caption{The plot of Gibbs free energy $(G)$ with horizon radius $(r_+)$ of the black string.} 
\label{fig:Sc4}
\end{figure}

The Gibbs free energy is also a substantial quantity for the thermodynamical system. The definition of  Gibbs free energy 
is given by $G=F+PV$. 
From the volume (\ref{Eq9}), the Helmholtz free  energy (\ref{III4}) and pressure (\ref{III5}), the  Gibbs free energy for the non-equilibrium system is calculated as
\begin{equation}
\label{III6}
G= \frac{\beta \alpha^2 r_+}{2 \pi}\left[\ln \left(\frac{9\alpha^5 r_{+}^4}{32 \pi}\right)^3 -6\right].
\end{equation}
The comparison of the equilibrium and non-equilibrium black string is depicted in Fig. \ref{fig:Sc4}. 
 Here, we see a significantly different behaviour of Gibbs free energy for the black string in equilibrium and non-equilibrium states. The Gibbs free energy without including thermal correction is zero. This means that no more work can be done. 
In contrast to the equilibrium case, the Gibbs free energy
under thermal fluctuation depends on the horizon radius and becomes more negatively valued for small black strings. {This suggests that the thermal fluctuation makes the reaction random or spontaneous.}

\section{Stability and compressibility}\label{sec4}
With the help of specific heat,  the  
stability and the phase transition during Hawking
evaporation process can be analysed.
To study the stability of the black string, we follow the signature of (specific) heat capacity. The positive sign of  heat capacity indicates the  black string is in a stable phase; however, the negative sign of heat capacity
shows an unstable phase transition.
Here, we calculate the heat capacity from the following definition: $C=-\left(\frac{\partial E}{\partial T}\right)$. 
For the given expressions (\ref{Eq6}) and (\ref{III3}),
the heat capacity of black string specifies to
\begin{equation}
C=-\left(\frac{\partial E}{\partial T}\right)=\pi \alpha r_{+}^2-4\beta.
\end{equation}
Here, we can see that the heat capacity takes a positive value ($C>0$) 
except $r_+^2<4\beta/\pi \alpha$. However, the heat capacity is always positive for the vanishing value of $\beta$  (no thermal correction) and shows a parabolic curve. The black string's phase transition from an unstable to a stable state occurs when the horizon radius increases. 

\begin{figure}[hbt]
\centering
\includegraphics[width=300pt]{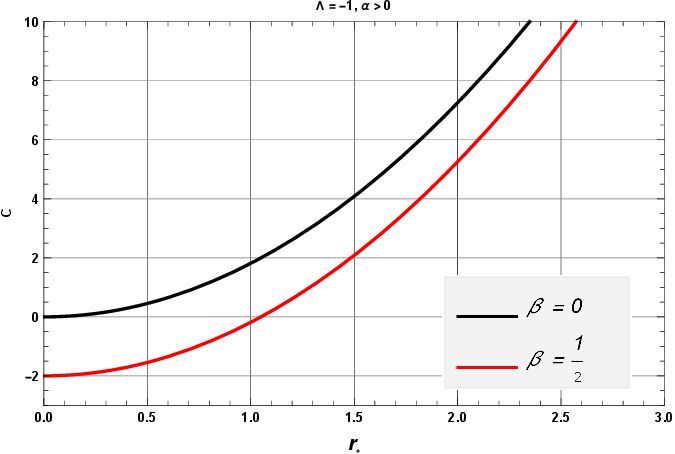} 
\caption{The plot of specific heat $(C)$ with horizon radius $(r_+)$ of the black string.} 
\label{fig:Sc5}
\end{figure}

To specify the region of stability and instability,  we plot the heat capacity concerning $r_+$ as depicted in Fig. 
 \ref{fig:Sc5}.
The figure shows that the fluctuation causes instability for a small radius of the black string in the region $r_+^2<4\beta/\pi \alpha$. However, no discontinuity is found in the black string's heat capacity. 

We considered the black string as a fluid system, so it is essential to derive the 
 compressibility, $K$, of the black string.
 This can be calculated with the help of   equations  (\ref{Eq9}) and (\ref{III5}) as
\begin{equation}
K= -\frac{1}{V}\left(\frac{\partial V}{\partial P}\right)=\frac{4 \pi^2 r_{+}^2}{\beta \alpha}\left[2-\ln \left(\frac{9 \alpha^5 r_{+}^4}{32 \pi}\right)\right]^{-1}.
\end{equation}
Here, we see that the compressibility of the black string depends on the fluctuation parameters inversely.

\section{Final remarks}\label{sec5}
It is possible to extend the spherical black hole solution to a cylindrical black hole (black string) solution \cite{13}. Considering the black string solutions in asymptotically AdS space, the CFT dual to such solutions can also be analysed \cite{23,24}. 
We have considered a system of a black string with a negative cosmological 
constant. Assuming the system is in equilibrium, we have discussed the thermodynamics of black string.
 
The small statistical fluctuation may play an essential role in the thermodynamics of the cylindrical black hole solutions with negative cosmological constant (black strings), so we have considered the respective modification to the entropy of the black string. The modified entropy eventually attributed to the thermodynamics of black string. We studied the change in the behaviour of the thermal equations of states due to thermal fluctuations. Here, we have found that the correction term of entropy of the black string becomes significant for small horizon black string.  The thermal fluctuation 
introduces more disorderedness to the system of small black strings.  
On the other hand, the fluctuation decreases the internal
 energy for the black string, which is significant for larger-sized black strings. However, the Helmholtz free energy becomes negative
 due to thermal fluctuation for a small black string. 
 The pressure is also calculated for the black string under thermal fluctuation in the extended phase space. The pressure 
 of the small black string becomes very high due to thermal fluctuation. 
We have calculated the Gibbs free energy  also for the system of black string under the influence of thermal fluctuation shows a matching behaviour  with the equilibrium counterpart
for the sufficiently large horizon radius.  

Finally, we have calculated the specific heat for the black string and found that the small statistical fluctuation causes instability for a small radius of the black string. However, excluding thermal fluctuation, the black string remains stable. 
The compressibility of the black string depends on the fluctuation parameters inversely.  
Since cylindrical black hole (black string) in $4D$ has a rich structure. Therefore, the present investigation can be used as a theoretical laboratory which may provide hints about the underlying feature of the interaction between geometry and quantum physics.
 Investigating the behaviour of $P-v$ criticality 
and quasi-normal modes for the black string solutions under the influence of small statistical fluctuation will also be interesting. 
This is the subject of future investigations.

\end{document}